\documentclass[12pt]{article}
\usepackage{graphicx}
\usepackage[caption=false]{subfig}
\newcommand\pubnumber{}
\newcommand\pubdate{}

\def\Title#1{\begin{center} {\Large #1 } \end{center}}
\def\Author#1{\begin{center}{ \sc #1} \end{center}}
\def\Address#1{\begin{center}{ \it #1} \end{center}}

\newcommand\pubblock{\rightline{\begin{tabular}{l} \pubnumber\\
         \pubdate  \end{tabular}}}
\newenvironment{Abstract}{\begin{quotation}  }{\end{quotation}}
\newenvironment{Presented}{\begin{quotation} \begin{center} 
             PRESENTED AT\end{center}\bigskip 
      \begin{center}\begin{large}}{\end{large}\end{center} \end{quotation}}
\def\Acknowledgements{\bigskip  \bigskip \begin{center} \begin{large}
             \bf ACKNOWLEDGEMENTS \end{large}\end{center}}
\def\beq{\begin{equation}}
\def\eeq#1{\label{#1}\end{equation}}
\def\eeqn{\end{equation}}

\def\beqa{\begin{eqnarray}}
\def\eeqa#1{\label{#1}\end{eqnarray}}
\def\eeqan{\end{eqnarray}}

\let\bar=\overbar

\def\M{{\cal M}}

\def\Dslash{\not{\hbox{\kern-4pt $D$}}}
\def\dslash{\not{\hbox{\kern-2pt $\del$}}}

\def\msb{{\bar{\ssstyle M \kern -1pt S}}}

\begin{document}
\begin{titlepage}
\pubblock

\vfill
\Title{Duality relations in proton diffraction dissociation and in DIS}.
\vfill
\Author{L\'aszl\'o Jenkovszky\footnote{Work was supported by the Program "Matter under Extreme Conditions" of the Nat. Ac. Sc. of Ukraine.}}
\Address{Bogolyubov Institute for Theoretical Physics (BITP),\\
	Ukrainian National Academy of Sciences \\14-b, Metrologicheskaya str.,
	Kiev, 03680, Ukraine;\\
	 jenk@bitp.kiev.ua}
\Author{Istv\'an Szanyi\footnote{Work was supported by HDFU (UMDSZ).}}
\Address{Uzhgorod National University, \\14, Universytets'ka str.,  
	Uzhgorod, 88000, Ukraine;\\
	 sz.istvan03@gmail.com}
\vfill
\begin{Abstract}
We use duality to relate resonances in missing mass $M$ to the large-mass diffraction dissociation of protons. In deep inelastic lepton-hadron scattering (DIS), hadronic resonances are related by duality in $Q^2$ to the low-$x$, smooth behaviour of the DIS structure functions. 
\end{Abstract}
\vfill
\begin{Presented}
Presented at EDS Blois 2017, Prague, \\ Czech Republic, June 26-30, 2017
\end{Presented}
\vfill
\end{titlepage}
\def\thefootnote{\fnsymbol{footnote}}
\setcounter{footnote}{0}

\section{Finite mass sum rule (FMSR) in proton diffraction dissociation}

Discovery \cite{DHSch} of finite energy sum rules (FESR) and duality between low-energy resonances and asymptotic Regge behaviour stimulated applications of duality in other areas of high-energy physics. 
Among these are finite-mass sum rules (FMSR) and parton-hadron duality in deep inelastic scattering (DIS). FMSR is an efficient tool in relating the contribution of resonances in the missing mass, $M$ produced in proton diffraction dissociation to the smooth large-$M$ Regge behaviour of the cross sections assumed in the triple-pomeron limit, based on the general optical theorem. 
 
In Refs. \cite{Magas} a method to include resonances in the missing mass $M$, based on a "reggeized Breit-Wigner" model following from duality was elaborated. Resonances in $M$ are generated by a direct-channel (pole) decomposition of the dual model. 
The single diffraction (SD) dissociation cross section in our model \cite{Magas} is :

\begin{equation}\label{SD}\frac{d^2\sigma_{SD}}{dtdM_i^2}={F_P}^2(t)F(x_B,t)\frac{\sigma_T^{Pp}(t,M_i^2)}{2m_p}(s/M_i^2)^{2(\alpha(t)-1)}\ln(s/M_i^2),\ i=1,2,\end{equation}
  \begin{equation}\label{total} \sigma_T^{Pp}(M_X^2,t)=Im\, A(M_X^2,t)=\frac{A_{N^*}}{\sum_n n-\alpha_{N^*}(M_X^2)}+Bg(t,M^2)=\end{equation}
  $$=A_{\it norm}\sum_{n=0,1,...}\frac{[f(t)]^{2(n+1)} Im\,\alpha(M^2_x)}{(2n+0.5-Re\, \alpha(M_X^2))^2+ (Im\,\alpha(M_X^2))^2}+Bg(t,M^2),$$
 where $\alpha(t)$ is the pomeron trajectory, $\alpha(M^2)$ is the Regge trajectory in the direct $\gamma-p$ channel, $Bg$ is the background and $F_p$ is the elastic $pPp$ vertex.
We use non-linear complex trajectories, typically 
\beq \alpha(x)=\alpha_0+\alpha_1 x+\alpha_2(\sqrt
{x_0}-\sqrt{x_0-x}),\ \ \ x=s,\ t,\M^2,
 \eeq{Eq:Tr}
 where $x_0$ is the lightest threshold in the given channel.

\begin{figure}[h]
\centering
\subfloat[\label{Res}]{%
	\includegraphics[height=6 cm]{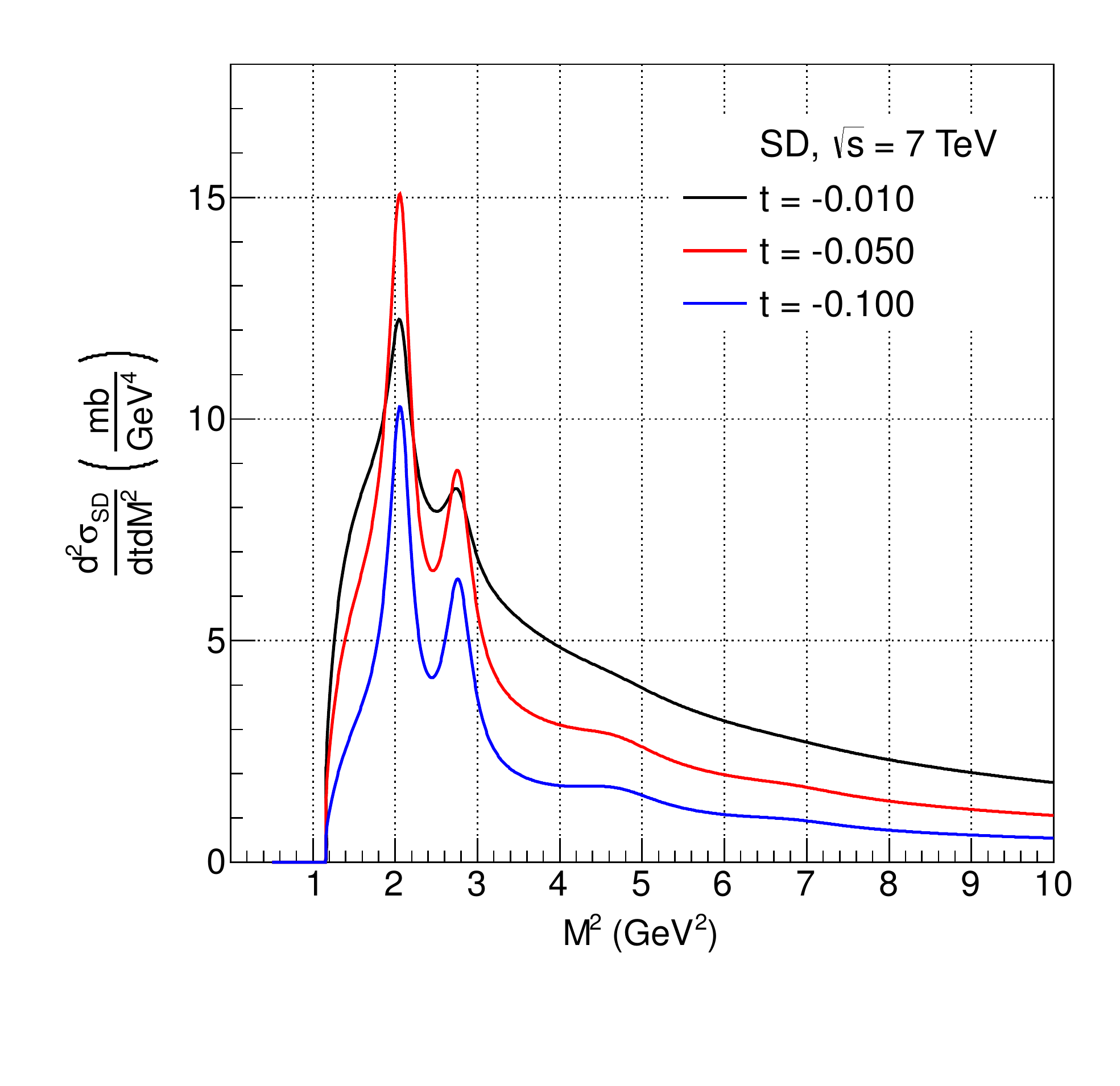}%
	}\hfil	
\subfloat[\label{FMSR}]{%
    \includegraphics[height=6 cm]{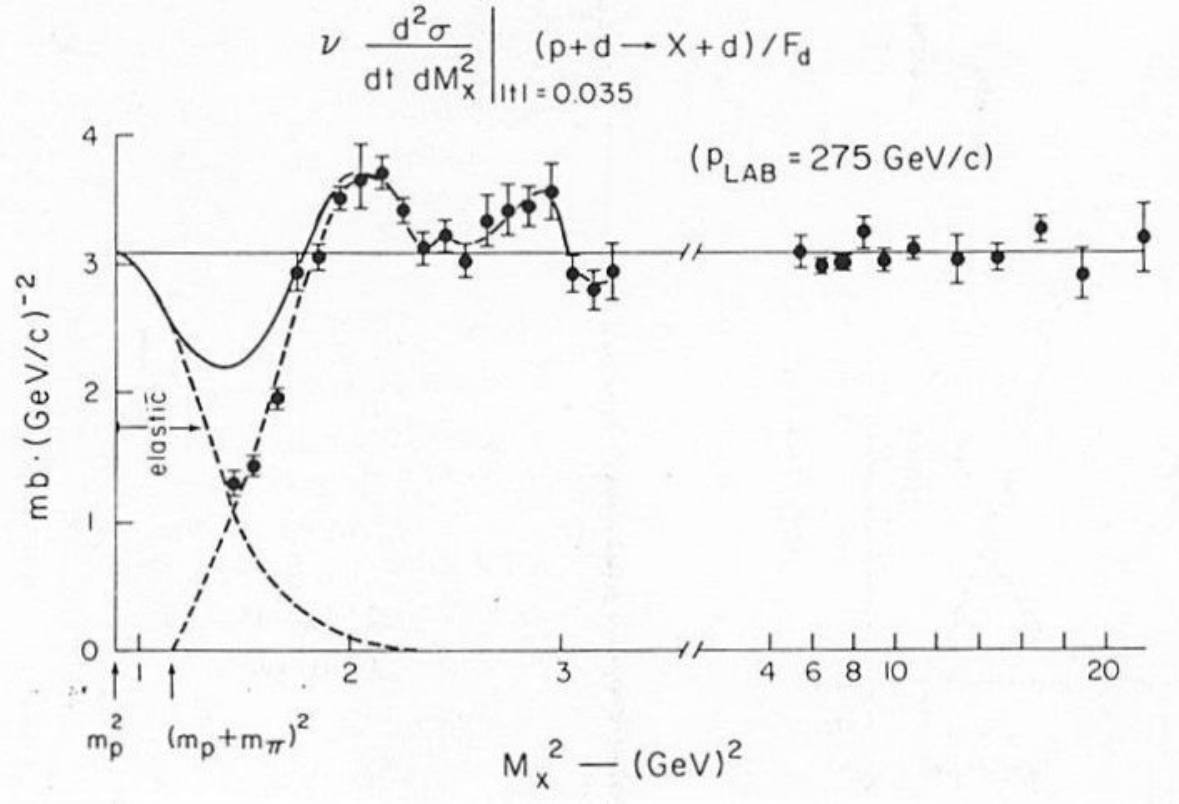}%
    }
\caption{(a) Double differential cross section of SD at the LHC for different values of $t$ calculated in \cite{Orava}. (b) A test of the FMSR \cite{Akimov}.} 
\end{figure}

The model reproduces the observable sequence of resonances in the missing mass  Fig. \ref{Res}, see also \cite{Schafer}.

The first moment FMSR 
$$
|t|\frac{d\sigma}{dt}+\int^{\nu_0}_0\nu\frac{d^2\sigma}{dtd\nu}=\int^{\nu_0}_0\nu\Bigl(\frac{d^2\sigma}{dtd\nu}\Bigr)_{Regge}d\nu
$$
states that the extrapolation of high $\nu$ behaviour of the function 
$\nu (d\sigma/dtd\nu)$ into the low $\nu$ region , where $\nu=M_x^2-M^2_p-t$  is the crossing-symmetric variable represents the average behaviour of the resonances and vice versa. FMSR were tested against single diffraction dissociation, see Fig.\ref{FMSR}. 

Duality sum rules work also in deep inelastic scattering (DIS), relating resonances, appearing at large $Q^2$ to smooth asymptotic behaviour at low $x$ (parton-hadron duality). 
Here again, resonances can be described by the above reggeized Breit-Wigner model, see \cite{Magas}.   

\section{Dual-Regge Structure Function and parton-hadron (Bloom-Gilman) duality}

\begin{figure}[htb]
\begin{center}
\includegraphics[height=4 cm,angle=0]{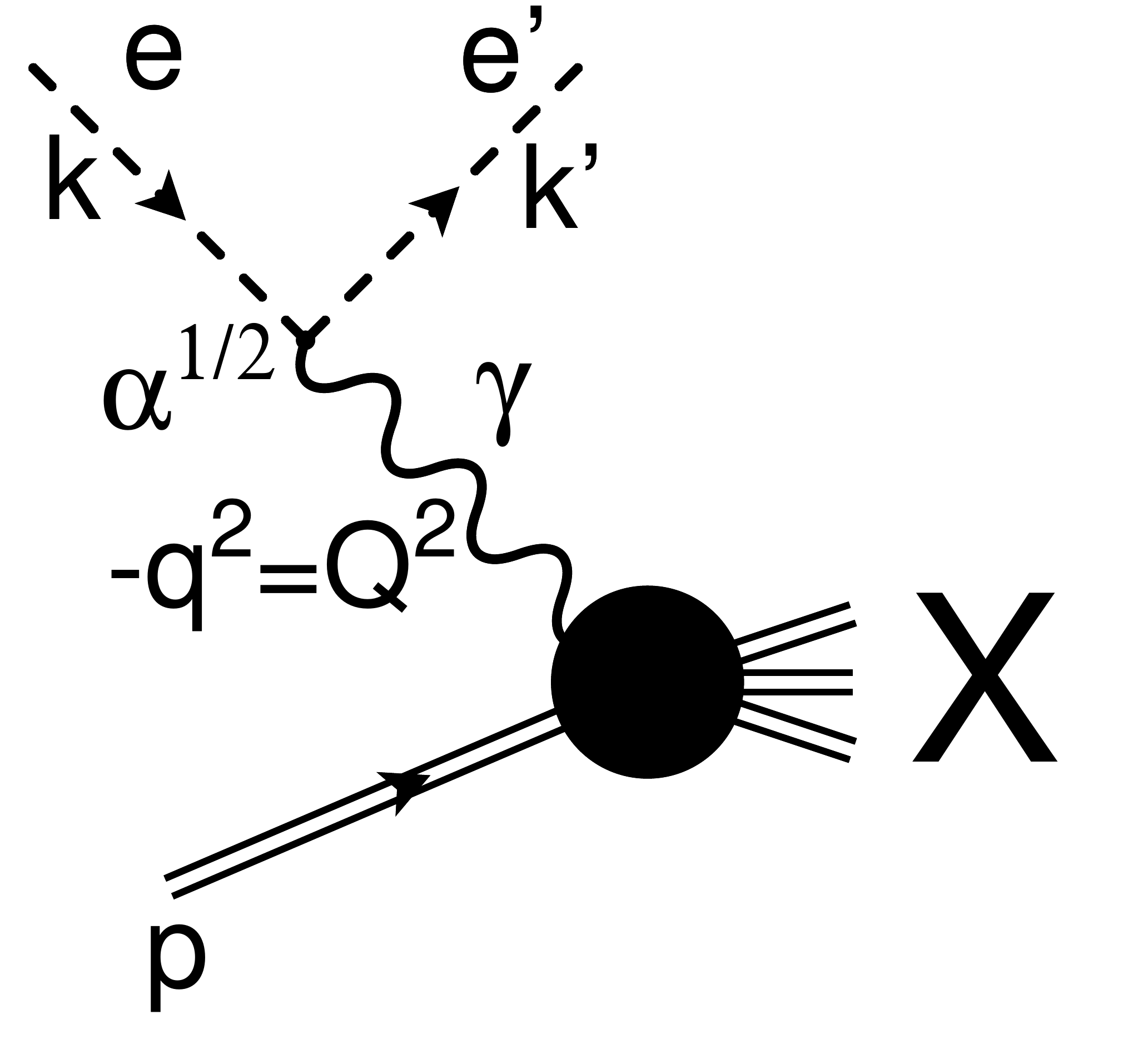}
\end{center}
\caption{Kinematic of deep inelastic scattering.}
\label{Diagram}
\end{figure}

{\it The kinematics of inclusive electron-nucleon scattering,
applicable to both high energies, typical of HERA, and low
energies as at JLab, is shown in Fig. \ref{Diagram}}.

The basic idea in our approach is the use the off-mass-shell
continuation of the dual amplitude with non-linear complex Regge
trajectories. We adopt the two-component picture of strong
interactions, according to which direct-channel resonances are
dual to cross-channel Regge exchanges and the smooth background in
the $s-$channel is dual to the pomeron exchange in the
$t-$channel. 

The cross section is related to the structure function by \beq
F_2(x,Q^2)={Q^2(1-x)\over{4\pi \alpha (1+4m^2 x^2/{Q^2})}}
\sigma_t^{\gamma^*p}~, \eeq{m23}, and use the norm where \beq
\sigma_t^{\gamma^*p}(s)={\cal I}m\  A(s,Q^2)~\eeq. The center of mass energy of the
$\gamma p$ system, the negative squared photon virtuality $Q^2$
and the Bjorken variable $x$ are related by \beq
s=W^2=Q^2(1-x)/x+m^2~. \eeq{m21}

In the Regge-dual approach with vector meson dominance implied,
Compton scattering can be viewed as an off-mass shell continuation
of a hadronic reaction, dominated in the resonance region by
non-strange ($N$ and $\Delta$) baryonic resonances. The scattering
amplitude can be written as a pole decomposition of the dual
amplitude and factorizes as a product of two vertices (form
factors) times the propagator: 
\beq \left[A(s,Q^2)\right]_{t=0} =
{\it N} \left\{ \sum_{r,n} {f_r^{2(n-n_r^{min}+1)}(Q^2) \over n -
\alpha_r(s)} + [A(s,Q^2)]_{t=0}^{BG}\right\}~, 
\eeq{dualampl}
where $\it N$ is an overall normalization coefficient, $r$ runs
over all trajectories allowed by quantum number conservation (in
our case $r=N^*_1,~N^*_2,~\Delta$) while $n$ runs from $n_r^{min}$
(spin of the first resonance) to $n_r^{max}$ (spin of the last
resonance - for more details see next section), and
$[A(s,Q^2)]_{t=0}^{BG}$ is the contribution from the background.
The functions $f_r(Q^2)$ and $\alpha_r(s)$ are respectively form
factors and Regge trajectory corresponding to the $r^{th}-$term.

\begin{figure}[htb]
\centering
\subfloat[\label{F_2}]{%
\includegraphics[height=13 cm]{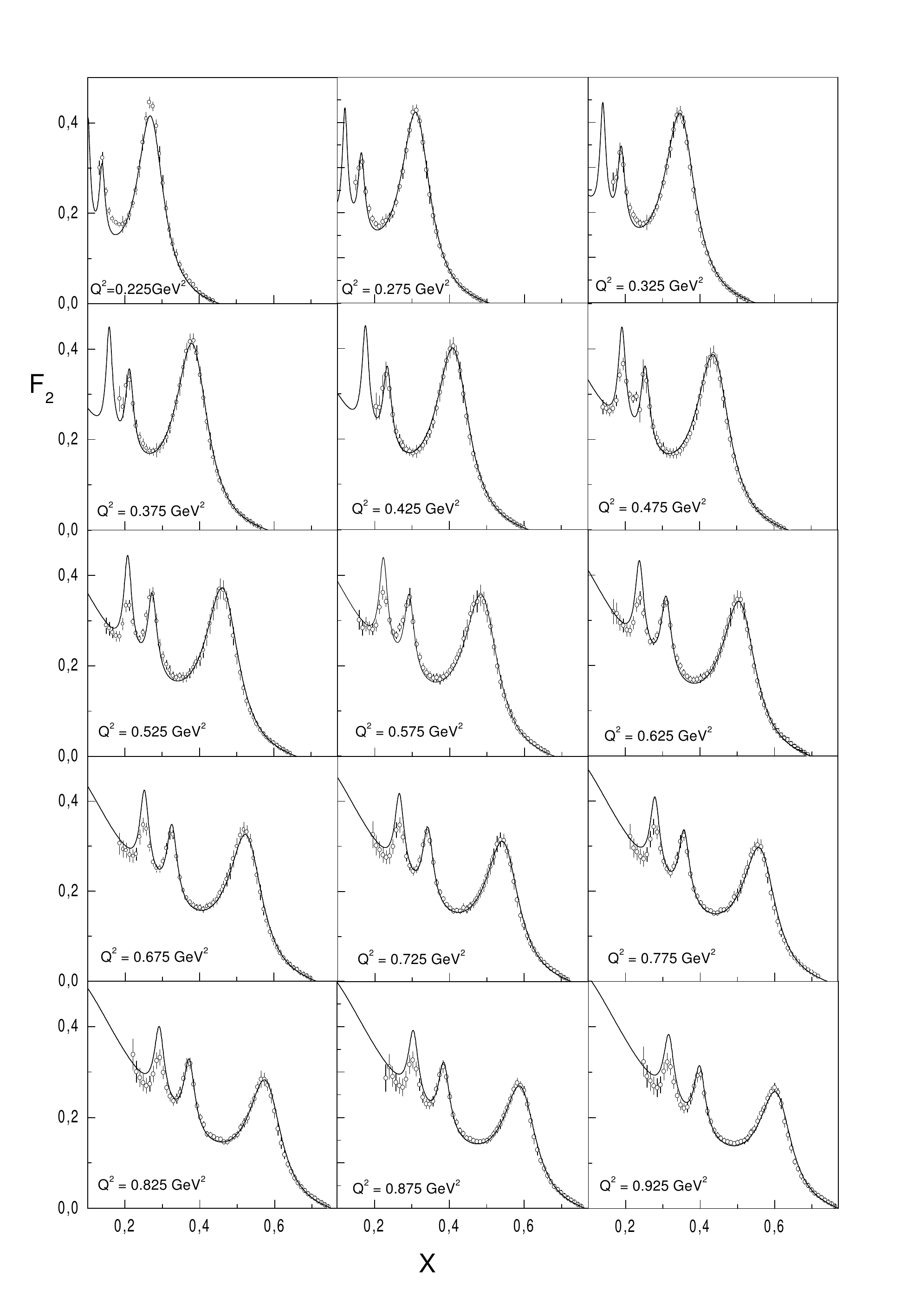}% 
}\par
\subfloat[\label{phtest}]{%
\includegraphics[height=5 cm]{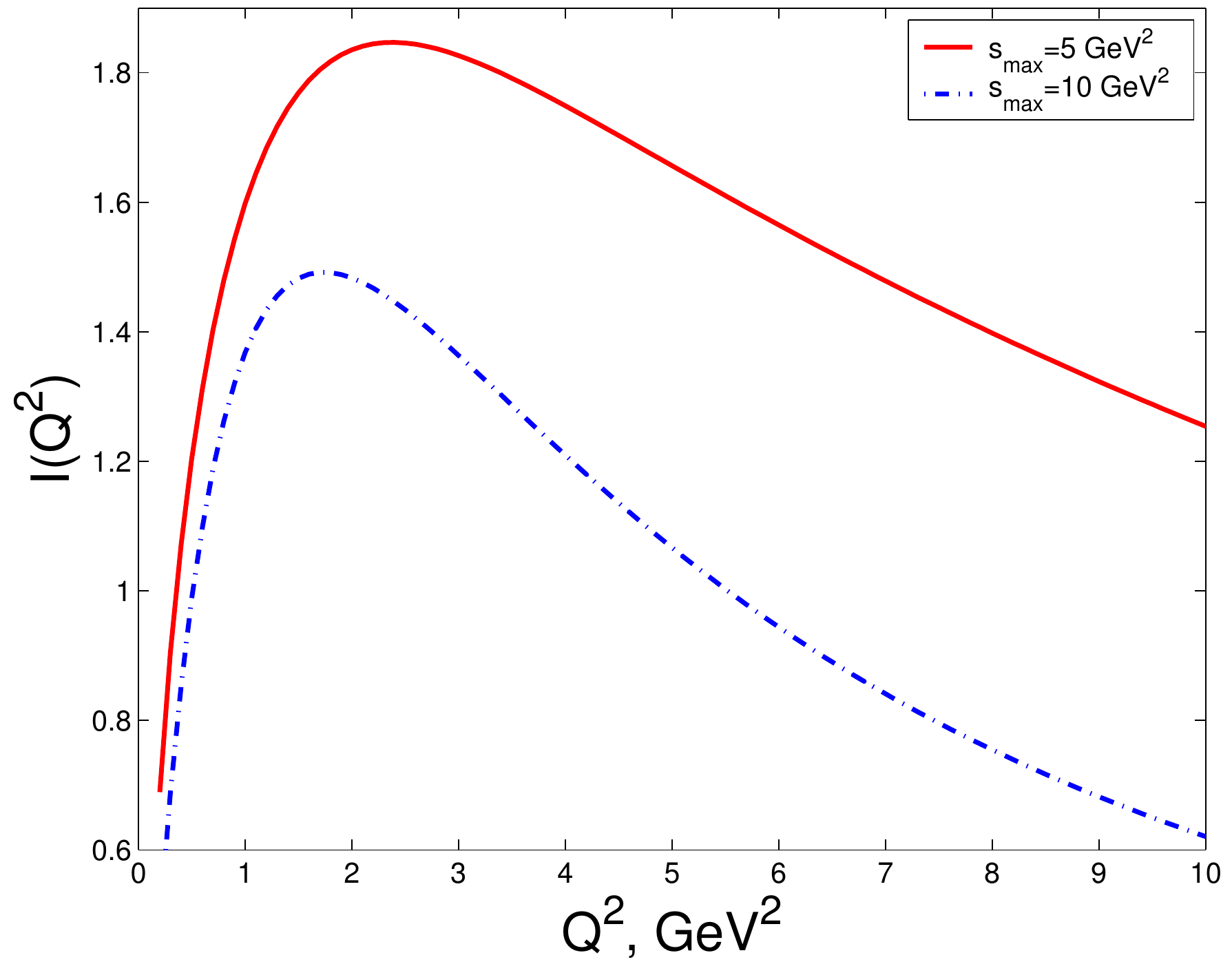}% 
}
\caption{(a) Model predictions \cite{Magas} for the structure function $F_2(x)$ at $Q^2=0.225\div 0.925$GeV$^2$. The data are from \cite{Osipenko}. (b) Global parton-hadron duality test for different values of $s_{max}$.}. 
\end{figure}

As seen from Fig. \ref{F_2}, the model fits almost perfectly the complicated resonance pattern,  however its rise towards small is $x$ too steep (see also \cite{Schafer}), which may be corrected by the use of asymptotically flatter Regge trajectories, namely: $\alpha(s)=\alpha_0-\sum_i\ln(1+\beta_i\sqrt{t_i-t}$ instead of (\ref{Eq:Tr}). 

%\section{Duality Relation}

Below we check the validity of parton-hadron duality
for our Regge-dual model by calculating the duality
relation 
\begin{equation}
I(Q^2)=\frac{I^{res.}}{I^{scale}},
\end{equation}
where
$$
I_{res.}(Q^2)=\int_{s_{min.}}^{s_{max}}dsF_2^{res.},
$$
$$
I_{scale}(Q^2)=\int_{s_{min.}}^{s_{max.}}dsF_2^{scale}
$$
using the model (\ref{dualampl}).

We fix the lower integration limit $s_{min}=s_0,$
varying the upper limit $s_{max}$ equal $5$ GeV $^2$ and $10$
GeV$^2$. These limits imply "global duality", i.e. a relation
averaged over some interval in $s$ (contrary to the so-called
"local duality", assumed to hold at each resonance position). For
fixed $Q^2$ the integration variable can be either $s$ (as in our
case), $x$ or any of its modifications ($x', \  \xi,...$) with
properly scaled integration limits. The difference may be
noticeable at small values of $Q^2$ due to the target mass
corrections (for details see e.g. \cite{Osipenko}).
These effects are typically non-perturbative and, apart from the
choice of the variables, depend on detail of the model.

The function $F_2^{Res}$ is our SF The results of the calculations for different
values of $s_{max}$ are shown in Fig. \ref{phtest}.

\Acknowledgements % if needed
We thank the Organizers for the inspiring atmosphere at the Meeting and for their support.

\end{document}